\documentclass[pra,amsmath,twocolumn,showpacs]{revtex4}
\usepackage{times}
\usepackage{graphicx}
\usepackage{amsmath}
\usepackage{mathptmx}

\newcommand{\intall}{\int_{-\infty}^{\infty}}

\newcommand{\bra}[1]{\langle#1|}
\newcommand{\ket}[1]{|#1\rangle}

\newcommand{\avg}[1]{\left\langle #1 \right\rangle}

\newcommand{\next}{\nonumber\\&\quad}
\newcommand{\etal}{\textit{et al.}}
\begin{document}

\title{On the Relationship between Resolution Enhancement 
and Multiphoton Absorption Rate in Quantum Lithography}
\author{Mankei Tsang}
\email{mankei@optics.caltech.edu}
\date{\today}
\affiliation{
Department of Electrical Engineering,
California Institute of Technology, Pasadena, California 91125, USA}
\begin{abstract}
  The proposal of quantum lithography [Boto \etal, \prl \textbf{85},
  2733 (2000)] is studied via a rigorous formalism. It is shown that,
  contrary to Boto \etal's heuristic claim, the multiphoton absorption
  rate of a ``NOON'' quantum state is actually lower than that of a
  classical state with otherwise identical parameters.  The
  proof-of-concept experiment of quantum lithography [D'Angelo \etal,
  \prl \textbf{87}, 013602 (2001)] is also analyzed in terms of the
  proposed formalism, and the experiment is shown to have a reduced
  multiphoton absorption rate in order to emulate quantum lithography
  accurately.  Finally, quantum lithography by the use of a jointly
  Gaussian quantum state of light is investigated to illustrate the
  trade-off between resolution enhancement and multiphoton absorption
  rate.
\end{abstract}
\pacs{42.50.Dv, 42.50.St}

\maketitle
\section{Introduction}
Optical lithography, the process in which spatial patterns are
transferred via optical waves to the surface of a substrate, has been
hugely successful in the fabrication of micro and nanoscale
structures, such as semiconductor circuits and microelectromechanical
systems. Conventional lithography cannot produce features much smaller
than the optical wavelength, due to the well known Rayleigh resolution
limit \cite{bornwolf}. As a result, beating the resolution limit for
lithography has become an important goal in the field of optics, with
far-reaching impact on other research areas including semiconductor
electronics and nanotechnology.

Among the many candidates proposed to supersede conventional
lithography, the use of extreme ultraviolet light for lithography
\cite{gwyn} has met numerous technical difficulties such as optics
imperfections and photoresist limitations.  Other proposals involve
multiphoton exposure \cite{yablonovitch,boto,agarwal,bentley,hemmer},
so that one can still use the more robust optics for long-wavelength
light, while obtaining some of the resolution improvements associated
with higher harmonics.  The feature size reduction, however, is not
nominal without taking care of the residual long-wavelength features
in a multiphoton absorption pattern.  Yablonovitch and Vrijen proposed
the use of several frequencies and narrowband two-photon absorption to
suppress such long-wavelength features \cite{yablonovitch}, while a
much more radical proposal by Boto \etal\ suggests the use of
$N$-photon quantum interference of $N$ spatially entangled photons
\cite{boto}. The so-called quantum lithography has several appeals,
such as arbitrary quantum interference patterns, generalization to
arbitrary number of photons, and the promise of multiphoton absorption
rate improvement via the use of entangled photons. Hence, despite
practical issues such as difficulties in generating a high dosage of
the requisite entangled photons and finding a suitable multiphoton
resist, interest in quantum lithography has been significant
\cite{boyd}.

Crucial to the future success of quantum lithography is the supposed
enhancement in the multiphoton absorption rate when entangled photons
are used. The absorption rate improvement for frequency anticorrelated
photons has been proved by Javanainen and Gould \cite{javanainen} and
Perina \etal\ \cite{perina}. Boto \etal\ further claimed that the
absorption rate should also improve for the spatially entangled
photons used in quantum lithography. This promise is absolutely vital
to the practicality of quantum lithography, because, as Boto \etal\
mentioned, classical multiphoton lithography is already infeasible for
large $N$, and quantum lithography would have been even worse if the
absorption rate was not enhanced, due to the much less efficient
generation of nonclassical light. Boto \textit{et al.}\ supported
their claim by arguing heuristically that the entangled photons are
constrained to arrive at the same place and at the same time. This
argument with respect to the spatial domain has, however, not been
substantiated with a more rigorous proof, and has been subject to
criticism \cite{steuernagel}. Unfortunately, the time domain treatment
\cite{javanainen,perina} cannot be directly carried over to the
spatial domain, because the former assumes a nearly resonant
multiphoton absorption process and does not require any temporal
resolution, but for quantum lithography the material response needs to
be spatially local to produce a high spatial resolution.

On the other hand, Boto \etal's formalism contains several crucial
approximations that remain to be justified. First, the photons are
implicitly assumed to arrive from a monochromatic source with a well
defined free-space wavelength $\lambda$, but the usual quantization
method of optical fields considers photons as quanta of
electromagnetic mode excitations, in which frequency appears only as a
dependent variable of the wave vector. It remains a question how
monochromatic optical fields as a boundary condition should be treated
in quantum optics. Second, they treat the two optical beams with
opposite transverse momenta as two discrete modes of optical fields,
but in free space, transverse momentum is a continuous variable, so
the discrete modes are evidently an approximation. Third, when
discussing the multiphoton absorption rate, Boto \etal\ regards
photons as objects in space and time that probabilistically arrive at
the photoresist, although it is well known that this interpretation of
photons is fundamentally flawed \cite{mandel}.

In this paper, starting from basic principles, I shall first
explicitly quantize the electromagnetic fields in
Sec.~\ref{quantization}, using assumptions consistent with Boto
\etal's proposal.  The formalism rigorously treats approximately
monochromatic optical fields as a boundary condition in the continuous
Fock space, and hence provides a theoretical underpinning to the
proposal of quantum lithography. The formalism also shows that,
regardless of the nonclassical spatial properties of the photons,
there exists an upper bound of multiphoton absorption rate, which
rules out any significant enhancement of multiphoton absorption rate
due to spatial effects only.  Next, using the developed formalism, I
shall analyze in Sec.~\ref{compare} the multiphoton absorption rate of
the so-called NOON state, and compare it with that of a classical
state with otherwise identical parameters. The analysis shows that,
despite both states having the same envelope for their interference
fringes, and despite the NOON state being able to reduce the
interference period of the fringes by a factor of $N$, the peak
multiphoton absorption rate of a NOON state is lower than that of a
classical state by a factor of $2^{N-1}$.  In Sec.~\ref{paraxial}, I
shall discuss the formalism in the paraxial regime, where it is
acceptable to regard photons as spatial objects described by a
configuration-space probability density. I shall then investigate the
proof-of-concept quantum lithography experiment by D'Angelo \etal\
\cite{dangelo} and show that the experiment requires a condition that
necessarily reduces the two-photon absorption rate, in order to
emulate quantum lithography accurately. Finally, I shall study the
multiphoton absorption of a jointly Gaussian multiphoton state, in
order to illustrate the trade-off between resolution enhancement and
multiphoton absorption rate.

\section{\label{quantization}
Quantization of Two-Dimensional, S-Polarized, Approximately
Monochromatic Electromagnetic Fields}
\subsection{Two Dimensional Approximation}
I shall start with the most general commutation relations for
creation and annihilation operators of continuous electromagnetic
field modes in free space \cite{mandel}:
\begin{align}
&\quad
[\hat{a}(k_x,k_y,k_z,s),\hat{a}^\dagger(k_x',k_y',k_z',s')]
\nonumber\\
&=\delta(k_x-k_x')\delta(k_y-k_y')\delta(k_z-k_z')\delta_{ss'},
\end{align}
where $k_x$, $k_y$, and $k_z$ are the independent continuous variables
for each mode of the electromagnetic fields, and $s$ denotes one of the
two polarizations perpendicular to the wave vector. The dependent
variable in this case is frequency $\omega$, determined by the
dispersion relation
\begin{align}
\omega^2 &= c^2(k_x^2+k_y^2+k_z^2).
\end{align}
The electric field operator is then given by \cite{mandel}
\begin{align}
\hat{\mathbf{E}}(x,y,z,t)&= \hat{\mathbf{E}}^{(+)}(x,y,z,t)
+ h.c.,
\\
\hat{\mathbf{E}}^{(+)}(x,y,z,t) &=
\frac{i}{(2\pi)^{\frac{3}{2}}}\sum_s \int d^3k
\left(\frac{\hbar\omega}{2\epsilon_0}\right)^{\frac{1}{2}}
\next\times
\hat{a}(k_x,k_y,k_z,s)\mathbf{e}(k_x,k_y,k_z,s)
\next\times
\exp(ik_x x + ik_y y + ik_z z-i\omega t),
\end{align}
where $h.c.$ denotes Hermitian conjugate, and
$\mathbf{e}(k_x,k_y,k_z,s)$ is the unit polarization vector
corresponding to one of the two polarizations orthogonal to the wave
vector. In the following I shall consider only the modes that
propagate in the positive $z$ direction in the $z-x$ plane, and only
the $s$ polarization normal to the $z-x$ plane.  This is consistent
with Boto \etal's proposal, and equivalent to assuming $k_z>
0$, $k_y\approx 0$, and choosing one $s$ such that $\mathbf{e}
=\hat{\mathbf{y}}$. Following Blow \etal\ \cite{blow}, I shall make
the following substitution to neglect the $y$ dimension:
\begin{align}
\int dk_y  &\to \frac{2\pi}{L_y},
\\
\hat{a}(k_x,k_y,k_z,s) &\to \hat{a}(k_x,k_z)\sqrt{\frac{L_y}{2\pi}},
\\
[\hat{a}(k_x,k_z),\hat{a}^\dagger(k_x',k_z')] &=\delta(k_x-k_x')\delta(k_z-k_z'),
\end{align}
where $L_y$ is the normalization length scale in the $y$ dimension. 
The electric field is thus simplified to
\begin{align}
\hat{E}^{(+)}(x,z,t) &=
\frac{i}{2\pi\sqrt{L_y}} \intall dk_x \int_0^\infty dk_z
\left(\frac{\hbar\omega}{2\epsilon_0}\right)^{\frac{1}{2}}
\next\times
\hat{a}(k_x,k_z)\exp(ik_x x + ik_z z-i\omega t).
\end{align}

\subsection{Propagating Fields}
To conform to classical optics conventions, I shall make $k_x$ and
$\omega$ the independent variables and $k_z$ the dependent variable,
following the procedure of Yuen and Shapiro \cite{yuen}.  This
coordinate transformation yields
\begin{align}
dk_xdk_z &= \frac{\omega}{c^2k_z}dk_x d\omega,
\label{differential}
\\
\hat{E}^{(+)}(x,t,z) &=
\frac{i}{2\pi\sqrt{L_y}} \int_{-\omega/c}^{\omega/c} dk_x \int_0^\infty d\omega
\left(\frac{\hbar\omega}{2\epsilon_0}\right)^{\frac{1}{2}}
\frac{\omega}{c^2k_z}
\next\times
\hat{a}\left(k_x,k_z\right)\exp(ik_x x -i\omega t+ ik_z z),
\end{align}
where $k_z=\sqrt{\omega^2/c^2-k_x^2}$ is now the dependent variable.
Consider the commutation relation for $\hat{a}(k_x,k_z)$ in terms of
the new variables,
\begin{align}
[\hat{a}(k_x,k_z),\hat{a}^\dagger(k_x',k_z')] 
&=\delta(k_x-k_x')\delta(k_z-k_z')
\\
&=\frac{c^2k_z}{\omega}\delta(k_x-k_x')\delta(\omega-\omega'),
\end{align}
where the factor $c^2k_z/\omega$ comes from Eq.~(\ref{differential}).
A new annihilation operator in terms of $k_x$ and $\omega$ should therefore be
defined as \cite{yuen_error}
\begin{align}
\hat{a}(k_x,\omega) &=\left(\frac{\omega}{c^2k_z}\right)^{\frac{1}{2}}
\hat{a}(k_x,k_z),
\end{align}
so that the new operators have the desired commutator,
\begin{align}
[\hat{a}(k_x,\omega),\hat{a}^\dagger(k_x',\omega')] 
&=\delta(k_x-k_x')\delta(\omega-\omega').
\end{align}
Writing $k_x$ as $\kappa$ as a shorthand, the electric field becomes
\begin{align}
\hat{E}^{(+)}(x,t,z) &=
i\left(\frac{\hbar}{8\pi^2\epsilon_0 c^2 L_y}\right)^{\frac{1}{2}}
\int_{-\omega/c}^{\omega/c} d\kappa \int_0^\infty d\omega
\next\times
\frac{\omega}{(\omega^2/c^2-\kappa^2)^{1/4}}
\next\times
\hat{a}\left(\kappa,\omega\right)
\exp\left(i\kappa x -i\omega t+ik_zz\right),
\end{align}
which is now expressed in terms of $z$-propagating modes, with 
transverse momentum $\kappa$ and frequency $\omega$ as the independent
degrees of freedom.

\subsection{Monochromatic Approximation}
We have now obtained a formalism that treats $\kappa$ and $\omega$ as
independent degrees of freedom and corresponds to the experimental
situation of quantum lithography, where optical fields are considered
as propagating modes. The spatial quantum properties of the
propagating waves are thus independent of the temporal properties. In
order to study only the spatial effect of resolution enhancement on
the multiphoton absorption rate, separate from the temporal effects
studied by Javanainen and Gould \cite{javanainen} and Perina \etal\
\cite{perina}, I shall assume, in consistency with Boto \etal's
formalism, that the photons are all approximately monochromatic with
the same frequency $\omega$. Again, following the conventions of Blow
\etal\ \cite{blow},
\begin{align}
\int d\omega \to \frac{2\pi}{T},
\quad
\hat{a}(\kappa,\omega) \to \hat{a}(\kappa)\sqrt{\frac{T}{2\pi}},
\\
[\hat{a}(\kappa),\hat{a}^\dagger(\kappa')] = \delta(\kappa-\kappa'),
\label{newcommutator}
\end{align}
where $T$ is the normalization time scale. The electric field
envelope is then defined as
\begin{align}
\hat{E}^{(+)}(x,z,t) &\equiv
\hat{E}^{(+)}(x,z)\exp(-i\omega t),
\\
\hat{E}^{(+)}(x,z) &=
i\left(\frac{\eta}{2\pi}\right)^{\frac{1}{2}}\int_{-\omega/c}^{\omega/c} d\kappa 
\gamma(\kappa)
\hat{a}\left(\kappa\right)
\exp\left(i\kappa x+ik_z z\right),
\label{envelope}
\end{align}
where
\begin{align}
\eta &\equiv \frac{\hbar\omega}{2\epsilon_0 c L_y T},
\end{align}
the magnitude of which is on the order of the optical intensity per
unit length in $x$ for one photon. $\gamma(\kappa)$ is a geometric
factor,
\begin{align}
\gamma(\kappa) &= \frac{1}{(1-c^2\kappa^2/\omega^2)^{1/4}},
\label{geometric}
\end{align}
which arises owing to the invariance of the formalism with respect to rotation
in the $z-x$ plane. See Appendix \ref{g(k)} for a detailed discussion on
the physical significance of $\gamma(\kappa)$.

\subsection{Continuous Fock Space Representation}
With the commutator described by Eq.~(\ref{newcommutator}), and the
electric field envelope in terms of the photon annihilation operator
in Eq.~(\ref{envelope}), a rigorous quantization of two-dimensional,
$s$-polarized, approximately monochromatic optical fields has been
established. To account for all possible configurations of a Fock
state in terms of the continuous transverse momentum, I shall define
the following $N$-photon eigenstate \cite{mandel,schweber},
\begin{align}
\ket{\kappa_{1},...,\kappa_{N}} &=
\frac{1}{\sqrt{N!}}
\hat{a}^\dagger(\kappa_{1})...\hat{a}^\dagger(\kappa_{N})\ket{0}.
\end{align}
A momentum-space representation of a Fock state $\ket{N}$ can be written as
\cite{schweber}
\begin{align}
\phi(\kappa_1,...,\kappa_N) &\equiv
\bra{\kappa_{1},...,\kappa_{N}}N\rangle
\\
&=\frac{1}{\sqrt{N!}}
\bra{0}\hat{a}(\kappa_{1})...\hat{a}(\kappa_{N})
\ket{N},
\label{momentumamplitude}
\end{align}
and the Fock state can then be written in terms of this representation,
\begin{align}
\ket{N} &= \int_{-\omega/c}^{\omega/c} d\kappa_{1}...
\int_{-\omega/c}^{\omega/c}d\kappa_{N}
\ket{\kappa_{1},...,\kappa_{N}}
\bra{\kappa_{1},...,\kappa_{N}}N\rangle
\\
&=\int_{-\omega/c}^{\omega/c}d\kappa_1...\int_{-\omega/c}^{\omega/c}d\kappa_N
\phi(\kappa_{1},...,\kappa_{N})
\ket{\kappa_{1},...,\kappa_{N}}.
\label{superposition}
\end{align}
$\phi$ is hereby defined as the momentum-space multiphoton amplitude,
which describes the configurations of $\kappa$'s for $N$
photons. $\phi$ must evidently satisfy the normalization condition,
\begin{align}
\int_{-\omega/c}^{\omega/c} d\kappa_1...
\int_{-\omega/c}^{\omega/c}d\kappa_N
|\phi(\kappa_1,,...,\kappa_N)|^2
&= 1,
\label{norm}
\end{align}
and the boson symmetrization condition,
\begin{align}
\phi(...,\kappa_n,...,\kappa_m,...) &=
\phi(....,\kappa_m,...,\kappa_n,...)
\textrm{ for any }n,m.
\label{boson}
\end{align}

\subsection{$N$-Photon Measurements at the Observation Plane}
We shall now observe the photons at $z = 0$, define
$\hat{E}^{(+)}(x)\equiv\hat{E}^{(+)}(x,0) $, and a spatial
multiphoton amplitude $\psi(x_1,...,x_N)$ as
\begin{widetext}
\begin{align}
\bra{0}\hat{E}^{(+)}(x_1)...\hat{E}^{(+)}(x_N)\ket{N}
&\equiv \sqrt{N!}i^N\eta^\frac{N}{2}\psi(x_1,...,x_N),
\\
\psi(x_1,...,x_N)&=
\frac{1}{(2\pi)^{N/2}}\int_{-\omega/c}^{\omega/c} d\kappa_1\gamma(\kappa_1)
...\int_{-\omega/c}^{\omega/c} d\kappa_N\gamma(\kappa_N)
\next\times
\phi(\kappa_1,...,\kappa_N)
\exp\left(i\sum_{n=1}^N\kappa_nx_n\right).
\label{amplitude}
\end{align}
The physical significance of $\psi$ is that its magnitude
squared is proportional to the ideal $N$-photon coincidence rate,
\begin{align}
\avg{:\hat{I}(x_1)...\hat{I}(x_N):} &=
\bra{N}\hat{E}^{(-)}(x_1)...\hat{E}^{(-)}(x_N)
\hat{E}^{(+)}(x_1)...\hat{E}^{(+)}(x_N)\ket{N}
\\
&=\left|\bra{0}\hat{E}^{(+)}(x_1)...\hat{E}^{(+)}(x_N)\ket{N}\right|^2
\\
&=N!\eta^N|\psi(x_1,...,x_N)|^2,
\end{align}
where 
\begin{align}
\hat{I}(x) \equiv \hat{E}^{(-)}(x)\hat{E}^{(+)}(x)
\end{align}
is the optical intensity operator.  Classically, an ideal $N$-photon
absorption pattern is given by $I^N(x)$. In quantum optics,
the average $N$-photon absorption rate becomes
\begin{align}
\avg{:\hat{I}^N(x):} &= \avg{:\hat{I}(x)...\hat{I}(x):}
=N!\eta^N|\psi(x,...,x)|^2,
\label{absorption}
\end{align}
which is the $N$-photon coincidence rate evaluated at the same
position $x_1=...=x_N=x$.

\subsection{Upper Bound of $N$-Photon Absorption Rate for an $N$-Photon State}
With the formalism outlined above, it turns out that one can already
derive an upper bound for the $N$-photon absorption rate of an
$N$-photon Fock state, without knowing the specific form of $\phi$,
using Schwarz's inequality $|\avg{f|g}|^2\le\avg{f|f}\avg{g|g}$,
\begin{align}
|\psi(x_1,...,x_N)|^2
&=\left|\frac{1}{(2\pi)^{N/2}}\int_{-\omega/c}^{\omega/c} d\kappa_1\gamma(\kappa_1)
...\int_{-\omega/c}^{\omega/c} d\kappa_N\gamma(\kappa_N)
\phi(\kappa_1,...,\kappa_N)
\exp\left(i\sum_{n=1}^N\kappa_nx_n\right)\right|^2
\\
&\le \frac{1}{(2\pi)^N}\left[\int_{-\omega/c}^{\omega/c} d\kappa_1...
\int_{-\omega/c}^{\omega/c} d\kappa_N
|\phi(\kappa_1,...,\kappa_N)|^2\right]
\next\times
\left[\int_{-\omega/c}^{\omega/c} d\kappa_1...\int_{-\omega/c}^{\omega/c} d\kappa_N
\left|\prod_{n=1}^N\gamma(\kappa_n)\exp\left(i\kappa_nx_n\right)\right|^2\right]
\\
&\le \frac{1}{(2\pi)^N}
\left[\int_{-\omega/c}^{\omega/c} d\kappa |\gamma(\kappa)|^2\right]^N=
\left(\frac{\pi}{\lambda}\right)^N,
\end{align}
\end{widetext}
where $\lambda = 2\pi c/\omega$ is the free-space wavelength.  Hence,
the $N$-photon absorption rate has a upper bound,
\begin{align}
\avg{:\hat{I}^N(x):} = N!\eta^N
|\psi(x,...,x)|^2\le N!\left(\frac{\pi\eta}{\lambda}\right)^N.
\end{align}
Recall that $\eta$ is on the order of the one-photon optical intensity
per unit length in $x$. The upper bound shows that the best
multiphoton absorption rate, regardless of the form of $\phi$, is on
the order of $I_0^N$, where $I_0$ is the optical intensity of one
photon focused to a width $\lambda$. Although this upper bound is
derived for the simple case of two-dimensional monochromatic optical
fields focused in one dimension, one expects that the situation should
remain qualitatively similar when the $y$ dimension is also
considered, leading to a maximum absorption rate on the order of the
$I_0^N$, where $I_0$ becomes the intensity of a photon focused to an
area of $\lambda^2$.  The enhancement of the multiphoton absorption
rate using nonclassical spatial properties of photons, if any, is
therefore likely to be very limited, compared with the linear
dependence of the absorption rate on $I_0$ obtainable using
frequency-anticorrelated photons \cite{javanainen,perina}.  This is due to
the resolution limit in the spatial domain that limits the spatial
bandwidth of the optical fields, as well as the perfectly local spatial
response of $N$-photon absorption assumed in Eq.~(\ref{absorption}).

\section{\label{compare}Multiphoton Absorption Rate of Quantum Lithography}
The chief results of Sec.~\ref{quantization} applicable to quantum
lithography are the definition of a normlizable momentum-space
multiphoton amplitude, Eq.~(\ref{momentumamplitude}), which is able to
describe arbitrary configurations of quantized, approximately
monochromatic, two-dimensional, $s$-polarized optical fields
containing $N$ photons, the definition of a spatial multiphoton
amplitude, Eq.~(\ref{amplitude}), and the average $N$-photon
absorption rate, Eq.~(\ref{absorption}), in terms of the spatial
amplitude. In the following I shall use these results to calculate
the $N$-photon absorption rates for a NOON state and a classical state
with otherwise identical parameters.

\subsection{$N$-Photon Absorption of a NOON State}
\begin{figure}[htbp]
\centerline{\includegraphics[width=0.45\textwidth]{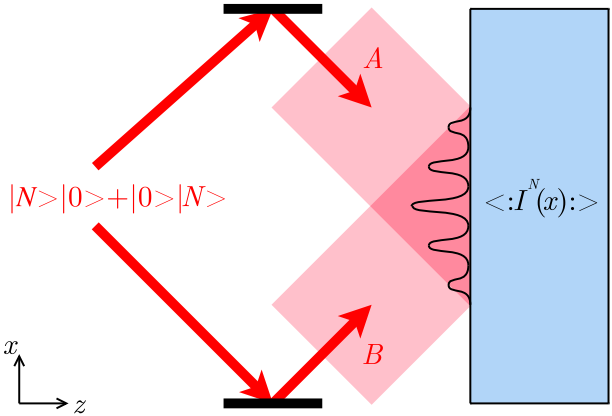}}
\caption{(Color online). Schematic of quantum lithography by the use of a NOON state.}
\label{lithography}
\end{figure}
In its simplest and most essential form, quantum lithography entails the
$N$-photon absorption of a NOON state \cite{boto} (Fig.~\ref{lithography}),
\begin{align}
\ket{NOON} &= \frac{1}{\sqrt{2}}\left(\ket{N}_A\ket{0}_B+\ket{0}_A\ket{N}_B\right)
\\
&=
\frac{1}{\sqrt{2N!}}\left[
(\hat{A}^\dagger)^N
+(\hat{B}^\dagger)^N\right]\ket{0},
\end{align}
where $A$ and $B$ label the two interfering beams
(Fig.~\ref{lithography}), and $\hat{A}^\dagger$ and $\hat{B}^\dagger$
are the creation operators for the two arms. In the continuous
momentum space, I shall express the corresponding annihilation
operators as
\begin{align}
\hat{A} &= \int d\kappa
\frac{1}{\sqrt{\Delta\kappa}}f\left(\frac{\kappa+\kappa_0}{\Delta\kappa}\right)
\hat{a}(\kappa),
\\
\hat{B} &= \int d\kappa
\frac{1}{\sqrt{\Delta\kappa}}
f\left(-\frac{\kappa-\kappa_0}{\Delta\kappa}\right)
\hat{a}(\kappa),
\end{align}
where $f(q)$ is a normalizable function of a dimensionless parameter $q$,
which satisfies $\int dq|f(q)|^2 = 1$ and describes the momentum
spread of modes $A$ and $B$. $\Delta\kappa$ is the momentum bandwidth,
and $\kappa_0$ is the tilt of the two arms.
$f((\kappa+\kappa_0)/\Delta\kappa)$ and
$f(-(\kappa-\kappa_0)/\Delta\kappa)$ are also assumed to be
orthogonal,
\begin{align}
\int d\kappa f\left(\frac{\kappa+\kappa_0}{\Delta\kappa}\right)
f^*\left(-\frac{\kappa-\kappa_0}{\Delta\kappa}\right) &= 0,
\end{align}
so that $\hat{A}$ and $\hat{B}$ satisfy the commutation relations
\begin{align}
[\hat{A},\hat{A}^\dagger] &=[\hat{B},\hat{B}^\dagger] = 1,
\quad
[\hat{A},\hat{B}^\dagger] = 0.
\end{align}
The momentum space amplitude, according to Eq.~(\ref{momentumamplitude}),
is
\begin{align}
\phi_{NOON}(\kappa_1,...,\kappa_N)
&= \frac{1}{\sqrt{N!}}\bra{0}\hat{a}(\kappa_{1})...\hat{a}(\kappa_{N})
\ket{NOON}
\\
&=\frac{1}{\sqrt{2\Delta \kappa^N}}
\Bigg[\prod_{n=1}^N f\left(\frac{\kappa_n+\kappa_0}{\Delta\kappa}\right)
\next
+\prod_{n=1}^N f\left(-\frac{\kappa_n-\kappa_0}{\Delta\kappa}\right)\Bigg].
\end{align}
One can check that this amplitude satisfies the normalization
condition, Eq.~(\ref{norm}). $\psi_{NOON}$ is thus determined to be
\begin{widetext}
\begin{align}
\psi_{NOON}(x_1,...,x_N) &= \frac{1}{\sqrt{2(2\pi\Delta\kappa)^N}}
\Bigg[\exp\left(-i\kappa_0\sum_{n=1}^Nx_n\right)
\prod_{n=1}^N\int d\kappa_n\gamma(\kappa_n-\kappa_0)
f\left(\frac{\kappa_n}{\Delta\kappa}\right)\exp(i\kappa_nx_n)
\nonumber\\&\quad+
\exp\left(i\kappa_0\sum_{n=1}^Nx_n\right)
\prod_{n=1}^N\int d\kappa_n\gamma(\kappa_n+\kappa_0)
f\left(-\frac{\kappa_n}{\Delta\kappa}\right)\exp(i\kappa_nx_n)\Bigg].
\end{align}
At $x_1=...=x_N=x$, $\psi_{NOON}$ becomes
\begin{align}
\psi_{NOON}(x,...,x) &= \frac{1}{\sqrt{2(2\pi\Delta\kappa)^N}}
\Bigg\{\exp\left(-iN\kappa_0x\right)
\left[\int d\kappa\gamma(\kappa-\kappa_0)
f\left(\frac{\kappa}{\Delta\kappa}\right)\exp(i\kappa x)\right]^N
\nonumber\\&\quad+
\exp\left(iN\kappa_0x\right)
\left[\int d\kappa\gamma(\kappa+\kappa_0)
f\left(-\frac{\kappa}{\Delta\kappa}\right)\exp(i\kappa x)\right]^N\Bigg\}.
\end{align}
\end{widetext}
I shall define a beam envelope function
\begin{align}
F(x) &\equiv \frac{1}{\sqrt{2\pi\Delta\kappa}}
\int d\kappa\gamma(\kappa-\kappa_0)
f\left(\frac{\kappa}{\Delta\kappa}\right)\exp(i\kappa x),
\end{align}
which is simply the electric field envelope of one of the optical
beams. We then have
\begin{align}
\psi_{NOON}(x,...,x) &=\frac{1}{\sqrt{2}}[F^N(x)\exp(-iN\kappa_0x)
\next
+F^N(-x)\exp(iN\kappa_0x)].
\end{align}
Assuming for simplicity an appropriate
$f(\kappa/\Delta\kappa)$ such that $F(x)$ is even, $\psi_{NOON}$ is
further simplified to
\begin{align}
\psi_{NOON}(x,...,x) &= \sqrt{2}
F^N(x)\cos(N\kappa_0 x),
\end{align}
and the $N$-photon absorption rate becomes
\begin{align}
\avg{:\hat{I}^N(x):}_{NOON} &= 2 N!\eta^N|F(x)|^{2N}
\cos^2(N\kappa_0 x).
\end{align}
The pattern consists of an envelope $|F(x)|^{2N}$ and an interference
fringe pattern $\cos^2(N\kappa_0 x)$. If $\Delta\kappa << \kappa_0$,
the envelope is much broader than each period of the interference
fringes, we then obtain the main result derived by Boto \textit{et
al.}, which is a multiphoton interference pattern $\cos^2(N\kappa_0x)$
with a period equal to $\pi/(N\kappa_0)$ and inversely proportional to
$N$.

\subsection{$N$-Photon Absorption of a Classical State}
\begin{figure}[htbp]
\centerline{\includegraphics[width=0.45\textwidth]{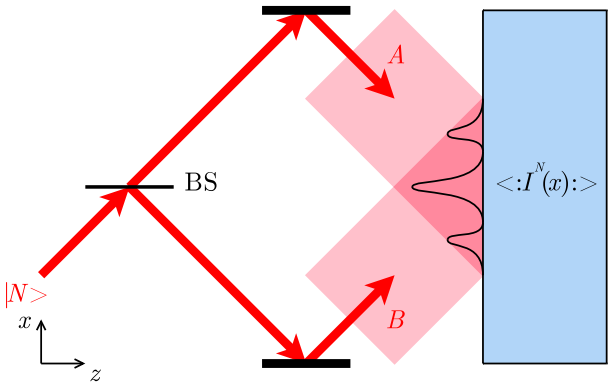}}
\caption{(Color online). Schematic of classical multiphoton lithography.}
\label{classical_lithography}
\end{figure}
The NOON state should be compared with a classical
$N$-photon state given by
\begin{align}
\ket{\Psi_C} &= \frac{1}{\sqrt{N!}}\left(
\frac{\hat{A}^\dagger+\hat{B}^\dagger}{\sqrt{2}}
\right)^N\ket{0},
\end{align}
which can be obtained, for example, by putting an $N$-photon state to
one of the inputs of a 50\%-50\% beam splitter
(Fig.~\ref{classical_lithography}). The momentum space amplitude is
\begin{align}
\phi_C(\kappa_1,...,\kappa_N) &= 
\prod_{n=1}^N\frac{1}{\sqrt{2}}\Bigg[
f\left(\frac{\kappa_n+\kappa_0}{\Delta\kappa}\right)
+f\left(-\frac{\kappa_n-\kappa_0}{\Delta\kappa}\right)\Bigg].
\end{align}
The amplitude is a product of one-photon amplitudes, underlining its
classical nature.  $\psi$ becomes
\begin{align}
\psi_C(x_1,...,x_N) &=
\prod_{n=1}^N \frac{1}{\sqrt{2}}
[F(x_n)\exp(-i\kappa_0x_n)
\next
+F(-x_n)\exp(i\kappa_0 x_n)].
\end{align}
Assuming again that $F(x)$ is even, the $N$-photon absorption rate is
\begin{align}
\avg{:\hat{I}^N(x):}_C &= 2^NN!\eta^N|F(x)|^{2N}\cos^{2N}(\kappa_0 x),
\end{align}
which has the same envelope $|F(x)|^{2N}$ as the NOON state, although
the interference period is fixed at $\pi/\kappa_0$.
The peak $N$-photon absorption rate is
\begin{align}
\avg{:\hat{I}^N(0):}_C &= 2^N N!\eta^N |F(0)|^{2N} =
2^{N-1}\avg{:\hat{I}^N(0):}_{NOON}.
\end{align}
Thus, even though we have meticulously carried out quantization and
normalization, we find that, under very general conditions, the peak
multiphoton absorption rate of a classical state is higher than that
of a NOON state by a factor of $2^{N-1}$, despite both having the same
envelope $|F(x)|^{2N}$. Hence, although the NOON state is able to
offer an $N$-fold enhancement of multiphoton interference resolution,
the NOON state manifestly does not have an enhanced multiphoton
absorption rate, as claimed by Boto \etal

\section{\label{paraxial}Paraxial Regime}
The factor $\gamma(\kappa)$ makes analytic calculations of the
multiphoton absorption pattern more difficult, and prevents an
intuitive understanding of the trade-off between resolution
enhancement and multiphoton absorption rate. To mitigate this issue,
in this section I shall work in the paraxial regime, where $\kappa \ll
\omega/c$, and $\gamma(\kappa)\approx 1$. This approximation
simplifies the analysis significantly and adequately describes most
optics experiments, including the proof-of-concept quantum lithography
experiment by D'Angelo \etal\ \cite{dangelo}.

\begin{figure}[htbp]
\centerline{\includegraphics[width=0.45\textwidth]{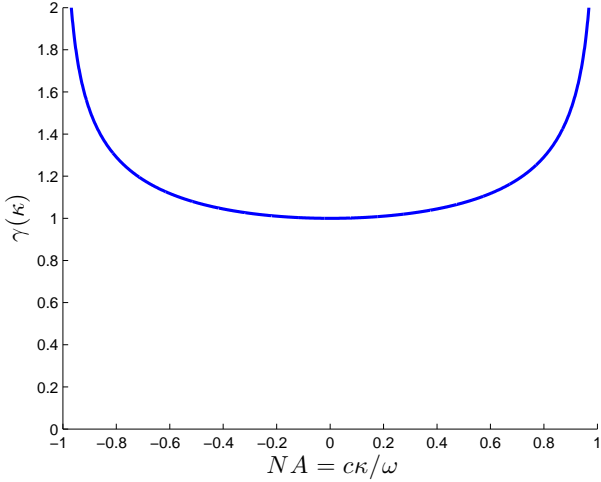}}
\caption{(Color online). A plot of $\gamma(\kappa)$ versus the
  numerical aperture $NA =c\kappa/\omega$.}
\label{geometric_plot}
\end{figure}

To justify the paraxial approximation, consider
Fig.~\ref{geometric_plot}, which plots $\gamma(\kappa)$ with respect
to the normalized parameter $NA = c\kappa/\omega$, defined as the
numerical aperture \cite{goodman}. One can see that $\gamma(\kappa)$
is relatively flat and $\approx 1$ for a wide range of $\kappa$. Even
for an $NA$ as high as $0.8$, $\gamma(\kappa)$ is only approximately
$1.3$, so in most practical cases, $\gamma(\kappa)$ provides only a
qualitatively unimportant correction factor.

In the paraxial regime, $\psi$ defined in Eq.~(\ref{amplitude}) 
becomes the familiar $N$-dimensional Fourier transform of $\phi$,
\begin{align}
\psi(x_1,...,x_N)
&\approx\frac{1}{(2\pi)^{N/2}}\intall d\kappa_1...
\intall d\kappa_N
\next\times
\phi(\kappa_1,...,\kappa_N)
\exp\left(i\sum_{n=1}^N\kappa_nx_n\right),
\label{fourier}
\end{align}
because $\kappa_n \ll \omega/c$ and $\gamma(\kappa_n)\approx 1$.
$\psi$ is then approximately normalized,
\begin{align}
\int dx_1...dx_N|\psi(x_1,...,x_N)|^2 &\approx 1.
\end{align}
$|\psi(x_1,...,x_N)|^2$ can thus be roughly regarded as the
configuration-space probability density of finding $N$ photons near
positions $x_1,...,x_N$ respectively. Provided that we do not localize
them too precisely, photons as particles in space are hence an
acceptable concept in the paraxial regime and described by a properly
normalized probability density.

The configuration-space model has been successfully applied to the
quantum theory of optical solitons \cite{lai}, where the slowly
varying temporal envelope approximation holds, so it is perhaps not
surprising that the model can also be applied to the spatial paraxial
domain, where the optical beam is relatively uniform.

\subsection{Simple Model of Multiphoton Absorption}
\begin{figure}[htbp]
\centerline{\includegraphics[width=0.3\textwidth]{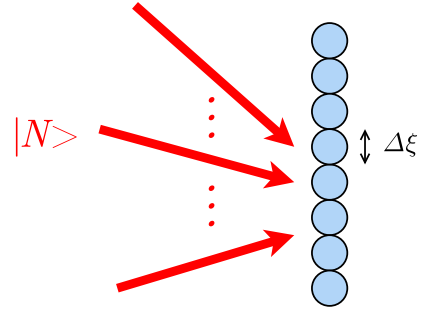}}
\caption{(Color online). A simple model of a multiphoton absorption
  material, consisting of many small individual multiphoton
  absorbers.}
\label{toy}
\end{figure}

This probabilistic spatial interpretation of photons also provides an
intuitive understanding of the expression for the multiphoton
absorption rate in Eq.~(\ref{absorption}). Consider a toy model for an
$N$-photon absorption material consisting of individual $N$-photon
absorbers, each occupying a width of $\Delta\xi$, as depicted in
Fig.~\ref{toy}.  The probability of all photons hitting the $m$th
absorber situated at $\xi_m$, thus exciting an $N$-photon absorption
event at this absorber, is given by
\begin{align}
P(\xi_m)\Delta\xi &= \int_{\xi_m-\Delta\xi/2}^{\xi_m+\Delta\xi/2} dx_1...
\int_{\xi_m-\Delta\xi/2}^{\xi_m+\Delta\xi/2} dx_N
|\psi(x_1,...,x_N)|^2,
\end{align}
where $P(\xi_m)$ is the probability density of the $N$-photon
absorption event.  The spatial resolution of the $N$-photon absorption
pattern evidently depends on $\Delta\xi$. To eliminate this dependence
and make the resolution depend solely on the resolution of the optical
fields, we shall make $\Delta\xi$ very small,
\begin{align}
P(\xi_m)\Delta\xi &\approx \Delta\xi^N |\psi(\xi_m,...,\xi_m)|^2,
\end{align}
so that in the limit of a continuous $N$-photon absorption material,
the probability density of $N$-photon absorption becomes
\begin{align}
P(x) &=\Delta\xi^{N-1}|\psi(x,...,x)|^2, 
\end{align}
which is proportional to $\avg{:\hat{I}^N(x):}$ given by
Eq.(\ref{absorption}). So, intuitively, an $N$-photon absorption event
occurs when all photons arrive within a very small neighborhood, and
the multiphoton absorption pattern is therefore approximately given by
the conditional probability distribution when all photons arrive at
the same place. This model has been used by Steuernagel to approximate
a four-photon absorption material by four discrete detectors
\cite{steuernagel}, although the explicit derivation here by the use
of a configuration space model confirms the intuition that a
multiphoton absorption event occurs when all photons arrive within a
small neighborhood.

It must be stressed that although the interpretation of
$|\psi(x_1,...,x_N)|^2$ as a configuration-space probability density
is only valid in the paraxial regime, the expression
Eq.~(\ref{absorption}) is always a valid description of an ideal
multiphoton absorption process, because of its dependence on the
optical intensity, a physically measurable quantity.

\subsection{\label{dangelo}Analysis of Proof-of-Concept Experiment
by D'Angelo \etal}
The proof-of-concept quantum lithography experiment by D'Angelo
\etal\ \cite{dangelo} remains well within the paraxial
regime, so an explicit analysis of the results can be carried out
relatively easily.  In this section, I shall show that the
coincidence rate detected by in D'Angelo \etal's
experiment is necessarily reduced, in order to emulate
quantum lithography accurately.

\begin{figure}[htbp]
\centerline{\includegraphics[width=0.45\textwidth]{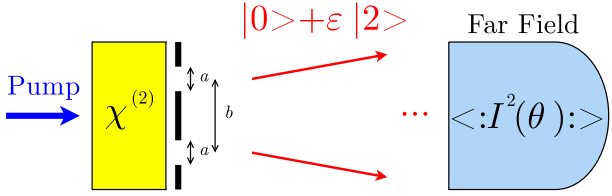}}
\caption{(Color online). Schematic of D'Angelo \etal's
proof-of-concept quantum lithography experiment \cite{dangelo}.}
\label{dangelo_setup}
\end{figure}
In the experiment depicted by Fig.~\ref{dangelo_setup}, the
spontaneously generated photon pair has the following quantum state,
\begin{align}
\ket{\Psi} &\approx \ket{0} + \epsilon\ket{2},
\end{align}
where $\epsilon$ depends on the efficiency of the spontaneous
parametric down conversion process, and must remain $\ll 1$, so that
the quantum state contains only zero or two photons in most cases.  In
D'Angelo \etal's analysis, the photon pair immediately
exiting the crystal is assumed to have perfect anti-correlation in
transverse momentum,
\begin{align}
\phi(\kappa_1,\kappa_2) &= \frac{1}{\sqrt{2}}
\bra{\kappa_1,\kappa_2}2\rangle
\sim\delta(\kappa_1+\kappa_2),
\label{anticorrelation}
\end{align}
as the pump beam is assumed to be relatively uniform across the
transverse plane of the crystal and the crystal is relatively short.
The spatial biphoton amplitude becomes
\begin{align}
\psi(x_1,x_2) &\sim \delta(x_1-x_2),
\label{spatialcorrelation}
\end{align}
and the photons are assumed to be perfectly correlated in space.
Clearly, both Eq.~(\ref{anticorrelation}) and
Eq.~(\ref{spatialcorrelation}) are approximations. In any case, we
shall first follow the approximate analysis and normalize the
expressions later. Immediately after exiting the crystal, the photon
pair passes through two slits of width $a$ spaced $b$
apart, resulting in a spatial amplitude
\begin{align}
\psi(x_1,x_2) &\sim 
\delta(x_1-x_2)
\next\times
\prod_{n=1}^2\Bigg[\textrm{rect}\left(\frac{x_n-b/2}{a}\right)
+\textrm{rect}\left(\frac{x_n+b/2}{a}\right)\Bigg]
\label{spatial}
\\
&\sim 
\delta(x_1-x_2)
\next\times\left[\textrm{rect}\left(\frac{x_1+x_2-b}{2a}\right)+
\textrm{rect}\left(\frac{x_1+x_2+b}{2a}\right)\right].
\label{spatialnoon}
\end{align}
Because the photons are assumed to be perfectly correlated in space,
they always pass through the same slit, resulting in a NOON state
\emph{in the spatial domain}. The momentum-space amplitude, on the
other hand, is given by
\begin{align}
\phi(\kappa_1,\kappa_2) &\sim
\textrm{sinc}\left[\frac{a(\kappa_1+\kappa_2)}{2}\right]
\cos\left[\frac{b(\kappa_1+\kappa_2)}{2}\right].
\label{farfield}
\end{align}
This is obviously not the NOON state in the momentum space for quantum
lithography.  To emulate quantum lithography indirectly, D'Angelo
\etal\ then let the photons propagate to the far field. Via Fraunhofer
diffraction \cite{goodman}, the angular two-photon coincidence
distribution, $\avg{:\hat{I}^2(\theta):}$ \cite{dangelo_note}, becomes
the magnitude squared of the Fourier transform of $\psi(x_1,x_2)$, or
\begin{align}
\avg{:\hat{I}^2(\theta):} &\propto
|\epsilon|^2\left|\phi\left(\frac{2\pi\theta}{\lambda},
\frac{2\pi\theta}{\lambda}\right)\right|^2
\\
&\sim |\epsilon|^2\textrm{sinc}^2\left(\frac{2\pi a\theta}{\lambda}\right)
\cos^2\left(\frac{2\pi b\theta}{\lambda}\right).
\end{align}
This expression is the same as that derived in
Ref.~\cite{dangelo}. $|\epsilon|^2$ is now regarded as the total
probability of two photons reaching the detection plane.  To be more
rigorous, however, $\psi(x_1,x_2)$ in Eq.~(\ref{spatialnoon}) and
$\phi(\kappa_1,\kappa_2)$ in Eq.~(\ref{farfield}) need to be
normalized. For example, the delta function in Eq.~(\ref{spatialnoon})
should be replaced by a sharp normalizable function,
\begin{align}
\delta(x_1-x_2) &\to \frac{1}{\sqrt{\alpha}}g\left(\frac{x_1-x_2}{\alpha}\right),
\label{replacedelta}
\end{align}
where $g(q)$ is a function of a dimensionless parameter $q$ and is
normalized according to $\int dq |g(q)|^2 = 1$. $\alpha$ is defined as
the biphoton coherence length, which depends on the phase matching
condition of the parametric down conversion process and the nonlinear
crystal length. $\alpha$ is assumed to be much smaller than $a$ and
$b$, but must still be non-zero in reality. The normalized $\psi$ then
becomes
\begin{align}
&\quad
\psi(x_1,x_2) 
\nonumber\\
&= \frac{1}{\sqrt{2\alpha a}}g\left(\frac{x_1-x_2}{\alpha}\right)
\next\times
\left[\textrm{rect}\left(\frac{x_1+x_2-b}{2a}\right)+
\textrm{rect}\left(\frac{x_1+x_2+b}{2a}\right)\right],
\end{align}
and the normalized $\phi$ becomes
\begin{align}
\phi(\kappa_1,\kappa_2) &= \frac{\sqrt{\alpha a}}{\pi}
G\left[\frac{\alpha(\kappa_1-\kappa_2)}{2}\right]
\next\times
\textrm{sinc}\left[\frac{a(\kappa_1+\kappa_2)}{2}\right]
\cos\left[\frac{b(\kappa_1+\kappa_2)}{2}\right],
\end{align}
where
\begin{align}
G(p) &\equiv \frac{1}{\sqrt{2\pi}}\int dq g(q)\exp(-ipq)
\end{align}
is the dimensionless Fourier transform of $g$.  The angular
distribution is hence
\begin{align}
\avg{:\hat{I}^2(\theta):} &\propto |\epsilon|^2\frac{\alpha
a}{\pi^2}|G(0)|^2\textrm{sinc}^2\left(\frac{2\pi a\theta}{\lambda}\right)
\cos^2\left(\frac{2\pi b\theta}{\lambda}\right),
\end{align}
which is proportional to $\alpha$. Thus, in order to produce a NOON
state in the near field and emulate quantum lithography accurately in
the far field, the photons need to pass through the same slit, the
biphoton coherence length needs to be small and is even assumed to be
zero in the analysis by D'Angelo \etal, but then the coincidence rate,
proportional to the biphoton coherence length, is necessarily reduced.

\subsection{Jointly Gaussian Multiphoton State}
In Sec.~\ref{compare}, we have studied the use of NOON state for
quantum lithography, and it has been shown that the NOON state has a
lower multiphoton absorption rate than a classical state. In
Sec.~\ref{dangelo}, we have also seen that, in order to approximate a
NOON state accurately in D'Angelo \etal's experiment, the multiphoton
absorption rate is necessarily reduced. While these results provide
evidence that it is probably impractical to use a NOON state for
multiphoton lithography, the NOON state is only one example of
infinitely many possible quantum states for optical fields, and other
quantum states might be able to perform better while still producing
an enhanced resolution.  For example, instead of producing enhanced
interference fringes with a minimum period on the order of
$\lambda/N$, Bj\"ork \etal\ \cite{bjork} considered another special
quantum state, called the reciprocal binomial state, in order to
produce a sharp interference spot, with a minimum width on the order
of $\lambda/N$, within a periodic pattern.  Still, it remains a
question whether this state can produce a significantly better
multiphoton absorption rate, as the NOON state is still a significant
component of the reciprocal binomial state. Steuernagel,
in particular, studied the four-photon reciprocal binomial state, with
four discrete detectors approximating an ideal four-photon absorption
material, and found that the multiphoton absorption rate is worse than
that of a classical state \cite{steuernagel}.

In this section, I shall study an arguably simpler and more intuitive
$N$-photon state that produces a quantum-enhanced Gaussian multiphoton
absorption spot, in the paraxial regime. I shall call this state a
jointly Gaussian state, which is a quantum generalization of the well
known classical Gaussian beams and is able to account for quantum
correlations of the photons. It is shown that, in certain limits, the
jointly Gaussian state is also able to reduce the size of the
multiphoton absorption spot by a factor of $N$ compared with the one
photon case, but the reduction of size is always accompanied by a
reduced multiphoton absorption rate.  For the jointly Gaussian state,
the quantum correlations of the photons, the size of the multiphoton
absorption spot, as well as the absorption rate can all be adjusted by
changing just two parameters, so a study of this state is able to
quantify and elucidate the trade-off between resolution enhancement
and multiphoton absorption rate.

\subsubsection{Many-Body Coordinate System}
Before defining a jointly Gaussian state, I shall first take a detour
and define a new many-body coordinate system, widely used in many-body
physics, which will significantly simplify the analysis later.
This coordinate transformation is
\begin{align}
K &= \frac{1}{N}\sum_{n=1}^N\kappa_n,
\quad
\kappa_n' =\kappa_n-K,
\end{align}
where $K$ is the average momentum, and $\kappa_n'$ is relative
momentum.  $K$ and $N-1$ of the $\kappa_n'$'s form a complete basis,
so I shall somewhat arbitrarily define $\kappa_N'$ as the extraneous
linearly dependent variable,
\begin{align}
\kappa_N' &=-\sum_{n=1}^{N-1}\kappa_n'.
\end{align}
The new differential is
\begin{align}
dKd\kappa_1'...d\kappa_{N-1}' &=\frac{1}{N}d\kappa_1...d\kappa_N, 
\end{align}
so a new multiphoton amplitude should be defined as
\begin{align}
\phi'(K,\kappa_1',...,\kappa_{N-1}') &= \sqrt{N}\phi(K+\kappa_1',...,K+\kappa_N'),
\label{newamplitude}
\end{align}
and normalized as
\begin{align}
\int dKd\kappa_1'...d\kappa_{N-1}'|\phi'(K,\kappa_1',...,\kappa_{N-1}')|^2
&= 1.
\end{align}
The multiphoton absorption rate in terms of the new amplitude in the paraxial
approximation is
\begin{align}
\avg{:\hat{I}^N(x):}
&=N!\eta^N \Bigg|
\sqrt{N}\int dKd\kappa_1'...d\kappa_{N-1}'
\next\times
\phi'(K,\kappa_1',...,\kappa_{N-1}')\exp\left(iNKx\right)\Bigg|^2.
\label{newabsorption}
\end{align}
The multiphoton absorption rate is therefore the magnitude squared of
the one-dimensional Fourier transform of $\phi'$, with respect to only
the total momentum $NK$.

\subsubsection{$N$-Photon Absorption of a Jointly Gaussian State}
In terms of the new coordinate system, we can now define the jointly
Gaussian state as follows,
\begin{align}
&\quad
\phi'(K,\kappa_1',...,\kappa_{N-1}')
\nonumber\\
&=\sqrt{C}
\exp\left(-\frac{K^2}{4B^2}\right)\exp\left(-\frac{1}{4\beta^2}
\sum_{n=1}^N\kappa_n'^2\right),
\label{defgauss}
\end{align}
where $B$ and $\beta$ are two parameters assumed to be real for
simplicity, $C$ is the normalization constant,
\begin{align}
C&=\left[\frac{N}{(2\pi)^N}\right]^{\frac{1}{2}}
\frac{1}{B\beta^{N-1}},
\label{normconst}
\end{align}
as derived in Appendix \ref{normalize}. This definition is inspired by
the well known jointly Gaussian distribution in statistics
\cite{garcia}. The form of Eq.~(\ref{defgauss}) is much simpler than a
general jointly Gaussian distribution because of bosonic symmetry, as
discussed in Appendix \ref{symmgauss}. The momentum amplitude of the
state given by Eq.~(\ref{defgauss}) is also very close to that of a
soliton state \cite{lai}, so one can obtain a general jointly
Gaussian state approximately by adiabatic control of spatial solitons
\cite{fini}.

The covariances of $K$ and $\kappa_n'$ are calculated in Appendix
\ref{covariances},
\begin{align}
\avg{K^2} &= B^2, \quad
\avg{\kappa_n'^2} = \left(1-\frac{1}{N}\right)\beta^2,
\end{align}
so $B$ is a measure of the spread in the average momentum, and $\beta$
is a measure of the spread in the momentum relative to the average.

The variance of $\kappa_n$, the momentum of each photon in the
original coordinates, is also derived in Appendix \ref{classical} and
given by
\begin{align}
\avg{\kappa_n^2} = B^2+\left(1-\frac{1}{N}\right)\beta^2,
\end{align}
$\kappa_n$ must be smaller than $\omega/c$, otherwise $k_z$ would
become imaginary, and $\kappa_n$ must be much smaller than $\omega/c$
for the paraxial approximation to hold. Moreover, if an optical system
has a certain aperture, it would also limit the transverse spatial
frequency \cite{goodman}.  Hence the variance of $\kappa_n$,
$\avg{\kappa_n^2}$, must be limited, and there exists a trade-off
between $B$ and $\beta$.

The multiphoton absorption pattern of the jointly Gaussian state can
be determined using Eq.~(\ref{newabsorption}),
\begin{align}
\avg{:\hat{I}^N(x):} &= N!\eta^N \sqrt{N}
\left(\frac{2}{\pi}\right)^{\frac{N}{2}}B\beta^{N-1}
\exp\left(-2N^2B^2x^2\right).
\label{gausspattern}
\end{align}
The pattern is a Gaussian, with a root-mean-square width given by
\begin{align}
W &\equiv \left(
\frac{\int dx x^2\avg{:\hat{I}^N(x):}}
{\int dx \avg{:\hat{I}^N(x):}}\right)^{\frac{1}{2}}
=\frac{1}{4NB}.
\end{align}
First, consider the case in which the photons are uncorrelated, and
the classical Gaussian state in the original system of coordinates is
given by a product of one-photon Gaussian amplitudes,
\begin{align}
\phi_C(\kappa_1,...,\kappa_N) &\propto
\prod_{n=1}^N\exp\left(-\frac{\kappa_n^2}{4\avg{\kappa_n^2}}\right).
\end{align}
As shown in Appendix \ref{classical}, this corresponds to the jointly
Gaussian state when
\begin{align}
B^2 &= \frac{\beta^2}{N} = \frac{\avg{\kappa_n^2}}{N}.
\end{align}
The classical variance of the average momentum is equal to the
variance of each momentum, $\avg{\kappa_n^2}$, divided by $N$.  This
is consistent with the statistics of independent photons.  The
multiphoton absorption width becomes
\begin{align}
W_C &= \frac{1}{4\sqrt{N\avg{\kappa_n^2}}},
\label{classical_width}
\end{align}
where the subscript $C$ denotes the value for a classical state.
Equation (\ref{classical_width}) can be regarded as the standard
quantum limit, and is better than the one-photon case by a factor of
$\sqrt{N}$. On the other hand, the minimum width is obtained when we
maximize $B$ so that $B=\sqrt{\avg{\kappa_n^2}}$ and let $\beta$ be
zero,
\begin{align}
W_{min} &= \frac{1}{4N\sqrt{\avg{\kappa_n^2}}}.
\end{align}
The factor-of-$N$ enhancement compared with one-photon
absorption, or the factor-of-$\sqrt{N}$ enhancement compared
with classical $N$-photon absorption, can be regarded as the ultimate
quantum limit, and is consistent with other quantum enhancement schemes
\cite{giovannetti}. This enhancement, however, comes with a heavy
price. Equation (\ref{gausspattern}) shows that the multiphoton
absorption rate is proportional to $B\beta^{N-1}$, so while increasing
$B$ and reducing $\beta$ makes the Gaussian pattern sharper, the
reduction in $\beta$ also reduces the multiphoton absorption rate,
more so for large $N$.

To quantify this trade-off, we shall fix $\avg{\kappa_n^2}$ as a given
resource, and define a spot size reduction factor $r$ with respect to
the classical case,
\begin{align}
r &\equiv \frac{W_C}{W} = \sqrt{\frac{N}{\avg{\kappa_n^2}}}B,
\end{align}
so that $r = 1$ corresponds to the standard quantum limit,
and $r = \sqrt{N}$ corresponds to the ultimate quantum limit.
We shall also define a normalized peak absorption rate $R$ with
respect to the rate in the classical case,
\begin{align}
R &\equiv \frac{\avg{:\hat{I}^N(0):}}{\avg{:\hat{I}^N(0):}_C}
=r\left(\frac{N-r^2}{N-1}\right)^{\frac{N-1}{2}}.
\end{align}

\begin{figure}[htbp]
\centerline{\includegraphics[width=0.45\textwidth]{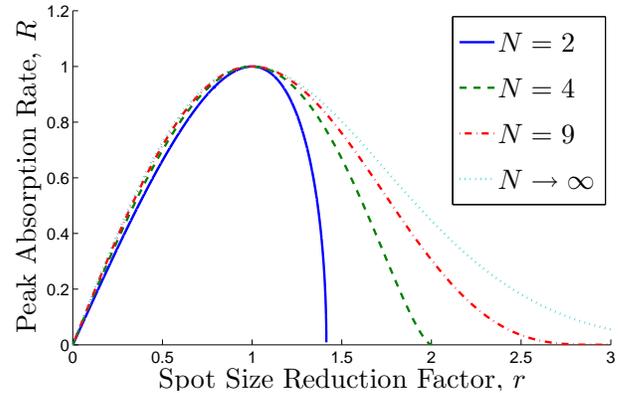}}
\caption{(Color online). Plots of peak multiphoton absorption rate
versus spot size reduction for several values of $N$. Both quantities
are normalized with respect to classical values. Interestingly, in the
limit of $N\to\infty$, $R \to r\exp[(1-r^2)/2]$.}
\label{peak_tradeoff}
\end{figure}

Figure \ref{peak_tradeoff} plots $R$ versus $r$ for several values of
$N$. This result is decidedly disappointing, as it shows that the
maximum multiphoton absorption rate is obtained when the state is a
classical state, or $r = 1$, and the peak rate monotonically decreases
to zero as the spot size is reduced. Furthermore, even if one is
willing to sacrifice the resolution and increase $\beta$,
the peak absorption rate is still reduced, because of its dependence
on $B$.

For applications such as multiphoton spectroscopy, spatial resolution
is not important, and it is more desirable to maximize the \emph{total}
multiphoton absorption rate. We can define the normalized total rate as
\begin{align}
R_{tot} &= \frac{\int dx \avg{:\hat{I}^N(x):}}{\int dx \avg{:\hat{I}^N(x):}_C}
=\left(\frac{N-r^2}{N-1}\right)^{\frac{N-1}{2}}.
\end{align}

\begin{figure}[htbp]
\centerline{\includegraphics[width=0.45\textwidth]{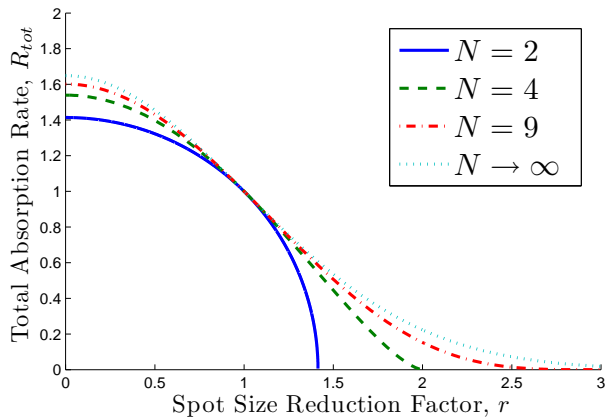}}
\caption{(Color online). Plots of total multiphoton absorption rate
versus spot size reduction for several values of $N$.}
\label{total_tradeoff}
\end{figure}

Figure \ref{total_tradeoff} plots $R_{tot}$ versus $r$. It can be seen
that the total rate does increase when one increases the spot size,
but the rate enhancement is very moderate. In fact, in the limit of
$N\to\infty$, $R_{tot}$ approaches
\begin{align}
R_{tot} &\to \exp\left(\frac{1-r^2}{2}\right),
\end{align}
so the ultimate rate enhancement, when resolution is completely
sacrificed and $r = 0$, asymptotically approaches $\exp(0.5)\approx
1.65$ for large $N$. This small enhancement of multiphoton absorption
rate is not likely to be useful.

To understand the above results, it is helpful to consider the
position correlation of the photons, derived in Appendix \ref{config},
\begin{align}
\avg{x_nx_m} &= \frac{1}{4N}\left(\frac{1}{NB^2}-\frac{1}{\beta^2}\right),
\quad n \neq m.
\end{align}
To obtain an enhanced multiphoton resolution, the bandwidth of the
average momentum, $B$, must be increased, leading to a positive
correlation in the momenta and a \emph{negative correlation} in the
positions of the photons.  So the photons are actually less likely to
arrive near one another, leading to a lower probability of the photons
hitting the same absorber and therefore a correspondingly less
multiphoton absorption rate. On the other hand, if $B$ is reduced, the
position correlation becomes positive, and the photons are more likely
to arrive close to one another, leading to a slightly enhanced total
absorption rate. Ultimately, how close the photons can arrive with
respect to one another is still restricted by the resolution limit of
the optical system, so the rate enhancement is not significant.

It must be stressed again that the preceding argument, as well as Boto
\etal's heuristic argument about photons constrained to arrive at the
same place, are only applicable to the paraxial regime, where the
positions of photons are relatively well defined quantities. The
example of the jointly Gaussian state shows that, even in the paraxial
regime, Boto \textit{et al.}'s heuristic argument is not correct, and
the photons are actually less likely to ``arrive at the same place''
when the resolution is enhanced. Although this has been shown by
Steuernagel for the specific example of four-photon reciprocal
binomial state \cite{steuernagel}, the study of joint Gaussian state
here confirms this fact for an arbitrary number of photons.

\section{Discussion and Conclusion}
In summary, through a rigorous study of quantum lithography and the
NOON state, an investigation of the proof-of-concept experiment by
D'Angelo \etal \cite{dangelo}, and an analysis of a jointly Gaussian
state, I have been unable to find any evidence, as far as the spatial
domain is concerned, that supports the heuristic claim by Boto \etal\
\cite{boto}, namely that the photons would be ``constrained to arrive
at the same place'' and the multiphoton absorption rate would be
enhanced due to spatial effects.  On the contrary, all examples show
that the multiphoton absorption rate is actually reduced, more so for
larger $N$, when a quantum state is used to enhance the resolution.

Admittedly, there are several assumptions involved in the analysis,
the negligence of time domain effects in particular.  As Javanainen
and Gould \cite{javanainen} and Perina \etal\ \cite{perina} have
shown, frequency entanglement can enhance the multiphoton
absorption rate, but this enhancement is likely to be independent of
the detrimental spatial effect, which seems to be an unavoidable
penalty incurred by the resolution enhancement effect itself. That
said, it remains to be proved whether taking time domain into account
would be able to eliminate or reverse the detrimental spatial effect.

Moreover, only the NOON state and the jointly Gaussian state have been
studied in this paper, but the possibility of other exotic quantum
states being able to enhance the resolution while maintaining a
respectable multiphoton absorption rate cannot be ruled out. In fact,
alternative strategies have already been proposed to solve the low
exposure problem of quantum lithography. For example, Agarwal \etal\
have shown that strong nonclassical beams from a parametric amplifier
can also produce enhanced two-photon interference fringes
\cite{agarwal}, albeit with a background worse than a classical
multiple exposure technique \cite{bentley}.  Hemmer \etal\ also
proposed the use of a narrowband multiphoton absorption material and
classical light \cite{hemmer}, similar to Yablonovitch and Vrijen's
proposal \cite{yablonovitch}, to achieve the same resolution
enhancement as quantum lithography.

In conclusion, in light of the results set forth, the original quantum
lithography scheme is unlikely to be practical in the near future.
Nonetheless, it has inspired many ongoing research efforts on the
elusive goal of beating the optical resolution limit, and should
therefore remain an interesting theoretical concept.

Discussions with Demetri Psaltis, Robert W.\ Boyd, Jonathan P.\
Dowling, Bahaa E.\ A.\ Saleh, and Paul W.\ Kwiat are gratefully
acknowledged. This work is financially supported by DARPA and the
National Science Foundation through the Center for the Science and
Engineering of Materials (DMR-0520565).

\appendix

\section{\label{g(k)}Physical Significance of the Geometric Factor $\gamma(\kappa)$}
To understand why the factor $\gamma(\kappa)$ arises in
Eq.~(\ref{amplitude}) of the formalism, consider the electric field
envelope given by Eq.~(\ref{envelope}),
\begin{align}
\hat{E}^{(+)}(x,z) &\propto
\int_{-\omega/c}^{\omega/c} d\kappa 
\gamma(\kappa)\hat{a}(\kappa)
\exp\left(i\kappa x+ik_z z\right),
\end{align}
where $k_z$ is the dependent variable given by $k_z =
\sqrt{\omega^2/c^2-\kappa^2}$.  We shall leave the form of
$\gamma(\kappa)$ unspecified and derive it purely from the fact that
the formalism is invariant under a rotation in the $z-x$
plane. Imagine that the electric field profile is rotated
anticlockwise in the $z-x$ plane, where $z$ is the horizontal axis and
$x$ is the vertical axis, by an angle $\theta$. This is equivalent to
defining new coordinates as follows,
\begin{align}
x' &= x\cos\theta  + z\sin\theta,
\quad
z' = -x\sin\theta + z\cos\theta.
\end{align}
The envelope becomes
\begin{align}
&\quad
\hat{E}^{(+)}(x',z')
\nonumber\\
&\propto
\int_{-\omega/c}^{\omega/c} d\kappa 
\gamma(\kappa)
\hat{a}(\kappa)
\next\times
\exp\big[i\kappa \left(\cos\theta x'+\sin\theta z'\right)
+ik_z\left(-\sin\theta x'+\cos\theta z'\right)\big]
\\
&=\int_{-\omega/c}^{\omega/c} d\kappa 
\gamma(\kappa)\hat{a}(\kappa)
\next\times
\exp\left[i\left(\kappa\cos\theta
-k_z\sin\theta \right)x'
+i\left(\kappa\sin\theta+k_z\cos\theta\right)z'
\right].
\end{align}
If we define the momenta in the new coordinate system to be
\begin{align}
\kappa' &= \kappa\cos\theta - k_z\sin\theta,
\quad
k_z' = \kappa\sin\theta + k_z\cos\theta,
\end{align}
evidently the new definitions still satisfy the dispersion relation
\begin{align}
\kappa'^2 + k_z'^2 &= \kappa^2 + k_z^2 = \frac{\omega^2}{c^2}.
\end{align}
More crucially, the coordinate transformation yields
\begin{align}
d\kappa &= \frac{k_z}{k_z'}d\kappa',
\label{trans}
\end{align}
so that the electric field envelope becomes
\begin{align}
\hat{E}^{(+)}(x',z')
&\propto
\int_{-\omega/c}^{\omega/c} d\kappa'
\frac{k_z}{k_z'}
\gamma(\kappa)\hat{a}(\kappa)
\exp\left(i\kappa'x'+ik_z'z'\right). 
\label{compare1}
\end{align}
If the electric field is invariant to such a rotation, we should be
able to define a new momentum-space operator $\hat{a}(\kappa')$ such
that
\begin{align}
\hat{E}^{(+)}(x',z')
&\propto
\int_{-\omega/c}^{\omega/c} d\kappa'
\gamma(\kappa')\hat{a}'(\kappa')
\exp\left(i\kappa'x'+ik_z'z'\right),
\label{compare2}
\end{align}
with the commutator
\begin{align}
[\hat{a}'(\kappa'),\hat{a}'^\dagger(\kappa'')] &=
\delta(\kappa'-\kappa'').
\end{align}
Comparing Eq.~(\ref{compare1}) and Eq.~(\ref{compare2}), we have
\begin{align}
\hat{a}'(\kappa') &=\frac{k_z}{k_z'}
\frac{\gamma(\kappa)}{\gamma(\kappa')}\hat{a}(\kappa),
\label{newoperator}
\end{align}
but the coordinate transformation also restricts the relation between
$\hat{a}(\kappa)$ and $\hat{a}'(\kappa')$,
\begin{align}
\hat{a}(\kappa) &=\left(\frac{d\kappa'}{d\kappa}\right)^{\frac{1}{2}}
\hat{a}'(\kappa')=
\left(\frac{k_z'}{k_z}\right)^{\frac{1}{2}}\hat{a}'(\kappa').
\label{operatorrelation}
\end{align}
Combining Eq.~(\ref{newoperator}) and Eq.~(\ref{operatorrelation})
yields
\begin{align}
\frac{\gamma(\kappa)}{\gamma(\kappa')}=\left(\frac{k_z'}{k_z}\right)^{\frac{1}{2}}.
\label{ratio}
\end{align}
For Eq.~(\ref{ratio}) to hold for any rotation, $\gamma(\kappa)$ must depend on
$\kappa$ only according to the following,
\begin{align}
\gamma(\kappa) &= Ck_z^{-\frac{1}{2}} = \frac{C}{(\omega^2/c^2-\kappa^2)^{1/4}},
\label{geometric2}
\end{align}
where $C$ is an arbitrary constant. Equation (\ref{geometric2}) is
identical to Eq.~(\ref{geometric}) with $C = \sqrt{\omega/c}$. Hence,
the factor $\gamma(\kappa)$ arises purely due to the invariance of the
formalism with respect to rotation in the $z-x$ plane. Furthermore,
the general transformation rule for $\hat{a}(\kappa)$ with respect to
a rotation is given by Eq.~(\ref{newoperator}), or
\begin{align}
\hat{a}(\kappa) &=
\left(\frac{k_z'}{k_z}\right)^{\frac{1}{2}}
\hat{a}'(\kappa') \\
&=
\left(\frac{\kappa\sin\theta+k_z\cos\theta}{k_z}\right)^{\frac{1}{2}}
\hat{a}'(\kappa\cos\theta-k_z\sin\theta).
\end{align}

\section{\label{gaussian}Properties of the Jointly Gaussian State}
In this section I shall derive several properties of the
jointly gaussian state,
\begin{align}
&\quad
\phi'(K,\kappa_1',...,\kappa_{N-1}')
\nonumber\\
&= \sqrt{C}\exp\left(-\frac{K^2}{4B^2}\right)\exp\left(-\frac{1}{4\beta^2}
\sum_{n=1}^N\kappa_n'^2\right),
\label{defgauss2}
\end{align}
where $C$ is the normalization constant, $B$ and $\beta$ are real
parameters, and $\kappa_N'$ is given by $-\sum_{n=1}^{N-1}\kappa_n'$.

\subsection{\label{normalize}Normalization}
To calculate $C$, consider the normalization
\begin{align}
C\int dKd\kappa_1'...\kappa_{N-1}'
\exp\left(-\frac{K^2}{2B^2}\right)
\exp\left(-\frac{1}{2\beta^2}\sum_{n=1}^N\kappa_n'^2\right) &= 1,
\\
C\sqrt{2\pi}B
\int d\kappa_1'...d\kappa_{N-1}'
\exp\left(-\frac{1}{2\beta^2}\sum_{n=1}^N\kappa_n'^2\right) &= 1.
\label{integral}
\end{align}
As $\kappa_N'=-\sum_{n=1}^{N-1}\kappa_n'$,
\begin{align}
\sum_{n=1}^N\kappa_n'^2 &=\sum_{n=1}^{N-1}\kappa_n'^2+
\left(\sum_{n=1}^{N-1}\kappa_n'\right)^2
=\sum_{n=1}^{N-1}\sum_{m=1}^{N-1}\kappa_n' A_{nm} \kappa_m',
\end{align}
where $A_{nm} = \delta_{nm}+1$ and $|A_{nm}| = N$.  Using the
normalization of a jointly Gaussian function \cite{garcia},
\begin{align}
&\quad\int d\kappa_1'...d\kappa_{N-1}'
\exp\left(-\frac{1}{2\beta^2}
\sum_{n=1}^{N-1}\sum_{m=1}^{N-1}\kappa_n'A_{nm}\kappa_m'\right)
\nonumber\\ &=
(\sqrt{2\pi})^{N-1}\beta^{N-1}|A_{nm}|^{-\frac{1}{2}}\\
&=\frac{(\sqrt{2\pi})^{N-1}\beta^{N-1}}{\sqrt{N}}.
\label{intgauss}
\end{align}
Substituting Eq.~(\ref{intgauss}) into Eq.~(\ref{integral}) gives
Eq.~(\ref{normconst}).

\subsection{\label{symmgauss} Bosonic Symmetry}
I shall now show that Eq.~(\ref{defgauss2}) is a
consequence of enforcing bosonic symmetry on a general jointly
Gaussian function. In the original coordinate system,
$\phi$ can be determined from Eq.~(\ref{newamplitude}),
\begin{align}
\phi(\kappa_1,...,\kappa_N) &=\sqrt{\frac{C}{N}}
\exp\Bigg[-\frac{1}{4B^2}
\left(\frac{1}{N}\sum_{n=1}^N\kappa_n\right)^2
\next
-\frac{1}{4\beta^2}
\sum_{n=1}^N\left(\kappa_n-\frac{1}{N}\sum_{m=1}^N\kappa_m\right)^2\Bigg],
\label{orig}
\end{align}
which can be rewritten as
\begin{align}
\phi(\kappa_1,...,\kappa_N) &=\sqrt{\frac{C}{N}}
\exp\left(-\frac{1}{4}\sum_{n=1}^N\kappa_n B_{nm}\kappa_m\right).
\label{gengauss}
\end{align}
Equation (\ref{gengauss}) is a general jointly Gaussian function, but
because of boson symmetry of $\phi$ as prescribed by
Eq.~(\ref{boson}), $B_{nm}$ must have identical on-axis components, as
well as identical off-axis components. With some algebra, $B_{nm}$ can
be determined from Eq.~(\ref{orig}),
\begin{align}
B_{nn} &=\frac{1}{N^2B^2}+\left(1-\frac{1}{N}\right)\frac{1}{\beta^2},
\\
B_{nm} &= \frac{1}{N^2B^2} - \frac{1}{N\beta^2},\quad n\neq m.
\end{align}
Any $B_{nm}$ with identical on-axis components and identical off-axis
components can be specified using $B$ and $\beta$, so
Eq.~(\ref{defgauss2}) can specify any general jointly Gaussian
functions with bosonic symmetry.

\subsection{\label{covariances}Covariances}
The covariances of momentum variables should be determined from the
probability distribution
\begin{align}
&\quad
|\phi'(K,\kappa_1',...,\kappa_{N-1}')|^2
\nonumber\\
&=C\exp\left(-\frac{K^2}{2B^2}\right)\exp\left(-\frac{1}{2\beta^2}
\sum_{n=1}^{N-1}\sum_{m=1}^{N-1}\kappa_n'A_{nm}\kappa_m'\right),
\end{align}
where $A_{nm} = \delta_{nm}+1$ is defined in Appendix \ref{normalize}.
The variance of $K$ is simply given by
\begin{align}
\avg{K^2} &= B^2,
\end{align}
while the covariance matrix for $\kappa_n'$'s is given by
\begin{align}
\avg{\kappa_n'\kappa_m'} &= \beta^2 A_{nm}^{-1}.
\end{align}
Because $A_{nm}$ only has two parameters, its inverse, defined as
$C_{nm}\equiv A_{nm}^{-1}$, is easy to calculate and is given by 
\begin{align}
C_{nn} = 1-\frac{1}{N};
\quad
C_{nm} = -\frac{1}{N}, \quad n\neq m.
\end{align}
We thus obtain the covariances,
\begin{align}
\avg{\kappa_n'^2} &= \left(1-\frac{1}{N}\right)\beta^2;
\quad
\avg{\kappa_n'\kappa_m'} = -\frac{\beta^2}{N},
\quad n\neq m.
\end{align}

\subsection{\label{classical}Classical Gaussian State}
The covariances of $\kappa_n$ in the original coordinate system are
\begin{align}
\avg{\kappa_n^2} &= \avg{\left(K+\kappa_n'\right)^2}
=\avg{K^2} + \avg{\kappa_n'^2}
\\
&=B^2 + \left(1-\frac{1}{N}\right)\beta^2,
\\
\avg{\kappa_n\kappa_m} &= \avg{K^2} + \avg{\kappa_n'\kappa_m'}
=B^2 -\frac{\beta^2}{N}.
\end{align}
So the photons are uncorrelated when $B^2 = \beta^2/N$,
and $\phi(\kappa_1,...,\kappa_N)$ in Eq.~(\ref{gengauss}) can be
written as
\begin{align}
\phi_C(\kappa_1,...,\kappa_N) &=\sqrt{\frac{C}{N}}
\prod_{n=1}^N\exp\left(-\frac{\kappa_n^2}{4\beta^2}\right),
\end{align}
a product of one-photon Gaussian amplitudes, and therefore a classical
state.

\subsection{\label{config}Configuration-Space Multiphoton Amplitude}
In the paraxial regime, the configuration-space multiphoton amplitude
can be obtained by Fourier transform of Eq.~(\ref{gengauss}),
\begin{align}
\psi(x_1,...,x_N) &\propto \exp\left(-\sum_{n,m}x_n B_{nm}^{-1}x_m\right),
\end{align}
which is determined using the well known characteristic function
of a jointly Gaussian distribution \cite{garcia}.
The configuration-space probability density is thus
\begin{align}
|\psi(x_1,...,x_N)|^2 &\propto \exp\left(-2\sum_{n,m}x_n B_{nm}^{-1}x_m\right),
\end{align}
and the covariance matrix for the photon positions is
$\avg{x_nx_m} = B_{nm}^{-1}/4$.
The position variance is then
\begin{align}
\avg{x_n^2} &= \frac{1}{4}\left[\frac{1}{N^2B^2}+
\left(1-\frac{1}{N}\right)\frac{1}{\beta^2}\right],
\end{align}
and the covariance is
\begin{align}
\avg{x_nx_m} &= \frac{1}{4N}\left(\frac{1}{NB^2}-\frac{1}{\beta^2}\right),
\quad n \neq m.
\end{align}


\begin{thebibliography}{}
\bibitem{yuen_error} Notice that Ref.~\cite{yuen} incorrectly assumes
$\hat{a}(k_x,k_y,\omega)=\hat{a}(k_x,k_y,k_z)$.
\bibitem{dangelo_note}In D'Angelo \etal's experiment, the two photons
are actually orthogonally polarized, and the measured quantity is
$\avg{:\hat{I}_1(\theta)\hat{I}_2(\theta):}$, where the subscripts $1$
and $2$ denote different polarizations. This measurement is the same
as the two-photon absorption pattern $\avg{:\hat{I}^2(\theta):}$ if
the two photons had the same polarization, so we shall neglect this
subtlety and use the expression $\avg{:\hat{I}^2(\theta):}$ for
simplicity.
\end{thebibliography}
\end{document}